\documentstyle[12pt,bezier]{article}
\begin{document}
\def \inbar{\vrule height1.5ex width.4pt depth0pt}
\def \xC{\relax\hbox{\kern.25em$\inbar\kern-.3em{\rm C}$}}
\def \xR{\relax{\rm I\kern-.18em R}}
\newcommand{\xZ}{Z \hspace{-.08in}Z}
\newcommand{\xbe}{\begin{equation}}
\newcommand{\xee}{\end{equation}}
\newcommand{\xbea}{\begin{eqnarray}}
\newcommand{\xeea}{\end{eqnarray}}
\newcommand{\xnn}{\nonumber}
\newcommand{\xkt}{\rangle}
\newcommand{\xbr}{\langle}
\newcommand{\xlll}{\left( }
\newcommand{\xrrr}{\right)}
\newcommand{\xcun}{\mbox{\footnotesize${\cal N}$}}
\title{Time-Dependent Diffeomorphisms as
Quantum Canonical  Transformations and the
Time-Dependent Harmonic Oscillator}
\author{Ali Mostafazadeh\thanks{E-mail address: 
amostafazadeh@ku.edu.tr}\\ \\
Department of Mathematics, Ko\c{c} University,\\
Istinye 80860, Istanbul, TURKEY}
\date{ }
\maketitle

\begin{abstract} 
Quantum canonical transformations corresponding to time-dependent 
diffeomorphisms of  the configuration space are studied. A special 
class of these transformations which correspond to time-dependent
dilatations is used to identify a previously unknown class of exactly
solvable time-dependent harmonic oscillators. The Caldirola-Kanai 
oscillator belongs to this class. For a general time-dependent harmonic
oscillator, it is shown that choosing the dilatation parameter to satisfy
the classical equation of motion, one obtains the solution of the
Schr\"odinger equation. A simple generalization of this result leads
to the reduction of the Schr\"odinger equation to a second order
ordinary differential equation whose special case is the auxiliary
equation of the Lewis-Riesenfeld invariant theory. Time-evolution 
operator is expressed in terms of a positive real solution of this
equation in a closed form, and the time-dependent position and 
momentum operators are calculated.
\end{abstract}
\vspace{3mm}
\vspace{3mm}

\baselineskip=24pt

\section{Introduction}
It is well-known that  in quantum mechanics the unitary transformations of 
the Hilbert space correspond to the canonical transformations of the 
classical mechanics. Unfortunately, these {\em quantum canonical 
transformations} are not as widely used as their classical counterparts.
The purpose of this article is to study the class of quantum canonical 
transformations defined by
	\xbe
	{\cal U}:=\exp\left[ \frac{i\epsilon(t)}{2}\{f(x),p\}\right]
	=\exp \left[i\epsilon(t)\sqrt{f(x)}\,p\,\sqrt{f(x)}\right]\;,
	\label{u}
	\xee
and demonstrate their utility in solving the Schr\"odinger equation for
time-dependent harmonic oscillators.\footnote{In Eq.~(\ref{u}), $x$ and
$p$ denote position and momentum operators, respectively, $\{~,~\}$ stands
for the anticommutator of two operators, and $f$ and $\epsilon$ are
arbitrary real-valued smooth functions.}

Time-dependent harmonic oscillators have been the subject of active
research since 1940's \cite{c-k}--\cite{pra}. This is because of the long list of
applications of this system in modelling a variety of physical phenomena.
Some recent applications of the time-dependent harmonic oscillators 
are in the study of the motion of ions in a Paul trap \cite{paul}, quantum 
mechanical description of highly cooled ions \cite{di-be-it-wi}, and 
emergence of nonclassical optical states of light due to a time-dependent 
dielectric constant \cite{ag-ku}.

An interesting property of time-dependent harmonic oscillators is that the
solution of the Schr\"odinger equation for this system can be reduced to that
of the corresponding classical equation of motion \cite{jpn}. It turns out 
that this reduction may be performed using a canonical transformation
of the form (\ref{u}). More generally, it is shown that the Schr\"odinger 
equation may be reduced to the solution of a certain second order ordinary
differential equation which involves a parameter $k^2$ with values
$-1,~0,$ and $1$. Any positive real solution of this equation with any
of the choices for $k^2$ yields the solution of the Schr\"odinger equation.
For $k^2=0$, one obtains the classical equation of motion. For $k^2=1$,
one finds the auxiliary equation of the invariant theory of Lewis and 
Riesenfeld \cite{le-ri}.

The organization of the paper is as follows. In Section~2 general results 
regarding time-dependent quantum canonical transformations (\ref{u})
and their special case corresponding to $f(x)=x$ are presented. These 
are then used in Section~3 to treat time-dependent harmonic oscillators.
Section~4 is devoted to the quantum canonical transformation which
leads to the reduction of the Schr\"odinger equation to the classical
equation of motion. The generalization of this result, its connection
with the invariant theory, and the calculation of the evolution operator
and the Heisenberg observables are also discussed in this section. 
Section~5 deals with the canonical transformations corresponding to
time-dependent diffeomorphisms which change the metric of the space. 
The conclusions are presented in Section~6.

\section{Time-dependent Quantum Diffeomorphisms and Dilatations}

Let us first recall the effect of a general time-dependent quantum canonical
transformation ${\cal U}={\cal U}(t)$ on the Hamiltonian $H=H(t)$ 
and the time-evolution operator $U=U(t)$, i.e., the relations
	\xbea
	H(t)\rightarrow H'(t)&=&
	{\cal U}(t)H(t)\,{\cal U}^\dagger(t)-i\,{\cal U}(t)\,
	\dot{\cal U}^\dagger(t)\;,
	\label{trans}\\
	U(t)\rightarrow U'(t)&=&{\cal U}(t)U(t)\,{\cal U}^\dagger(0)\;,
	\label{trans-u}
	\xeea
where a dot means a time-derivative and $\hbar$ is set to unity.
These equations are direct  consequences of the requirement that the
Schr\"odinger equation
	\xbe
	H(t)U(t)=i \,\dot U(t)\,,~~~~~~~U(0)=1\;,
	\label{sch-eq-u}
	\xee
must be preserved under the action of ${\cal U}$. Note that under a
time-dependent quantum canonical transformation the Hamiltonian
undergoes an affine (non-linear) transformation. Hence, unlike the
dynamics (the Schr\"odinger equation) the energy spectrum is
not preserved.

Next let us study the effect of the transformation induced by (\ref{u}).
In order to compute the transformed Hamiltonian $H'$, one must first
explore the effect of ${\cal U}$ on the position and momentum operators.
A rather lengthy calculation shows that
	\xbea
	x\to x'&:=&{\cal U}\:x\:{\cal U}^\dagger\:=\:{\cal F}_1(x)\;,
	\label{x}\\
	p\to p'&:=&{\cal U}\:p\:{\cal U}^\dagger\:=\:\frac{1}{2}\{{\cal F}_2(x),p\}
	\:=\:\sqrt{{\cal F}_2(x)}\:p\:\sqrt{{\cal F}_2(x)}\;,
	\label{p}
	\xeea
where
	\[ {\cal F}_1(x):=e^{\epsilon(t)f(x)\frac{d}{dx}}\,x\;,~~~~
	{\cal F}_2(x):=f(x)\,e^{\epsilon(t)f(x)\frac{d}{dx}}\,f^{-1}(x)\;.\]
In the derivation of these formulae use is made of
Baker-Campbell-Hausdorff  formula: 
	\[ e^ABe^{-A}=B+[A,B]+\frac{1}{2!}[A,[A,B]]+\cdots\;,\]
and the identities
	\xbea
	\left[\frac{i}{2}\{ f_1(x),p\},f_2(x)\right]&=&f_1(x)\frac{d}{dx}f_2(x)\;,
	\label{diff1}\\
	\left[ \frac{i}{2}\{ f_1(x),p\}, \frac{i}{2}\{f_2(x),p\}\right]&=&
	\frac{i}{2}\{f_3(x),p\}\;,~~~~
	f_3(x):=f_1\frac{d}{dx}f_2(x)-f_2\frac{d}{dx}f_1(x)\;,
	\label{diff2}
	\xeea
where $f_1$ and $f_2$ are arbitrary differentiable functions.

Eqs.~(\ref{diff1}) and~(\ref{diff2}) show that $\frac{1}{2}\{ f(x),p\}$
generate diffeomorphisms of $\xR$. Consequently, the
transformations (\ref{u}) are quantum canonical transformations associated
with time-dependent diffeomorphisms of  the configuration space.

In view of Eqs.~(\ref{trans}), (\ref{x}) and (\ref{p}), one has
	\xbe
	H'=H'(x',p';t)=H(x',p';t)-\frac{\dot\epsilon(t)}{2}\{f(x),p\}\;.
	\label{h'}
	\xee

Now let us concentrate on a subclass of quantum canonical
transformations of the form (\ref{u}) corresponding to the choice
$f(x)=x$. In this case, Eqs.~(\ref{x}), (\ref{p}), and (\ref{h'})
reduce to
	\xbea
	x\to x'&=&e^{\epsilon(t)}x\;,~~~~~
	p\to p'\:=\:e^{-\epsilon(t)}p\;,
	\label{trans-xp}\\
	H\to H'&=&H(x',p';t)-\frac{\dot\epsilon(t)}{2}\{x,p\}\;.
	\label{trans-h}
	\xeea
Hence this choice yields time-dependent dilatations of the configuration
space.

For a Hamiltonian of the standard form 
	\xbe
	H=\frac{p^2}{2m(t)}+V(x,t)\;,
	\label{natural}
	\xee
Eqs.~(\ref{trans-xp}) and (\ref{trans-h}) lead to
	\xbe
	H'=\frac{p^2}{2m(t)e^{2\epsilon(t)}}+V(e^{\epsilon(t)}x,t)
	-\frac{\dot\epsilon(t)}{2}\{x,p\}\;.
	\label{H'2}
	\xee
Therefore the transformed  Hamiltonian is not of the standard form 
(\ref{natural}). It can however be put in this form by the canonical 
transformation defined by
	\xbe
	{\cal U'}=\exp\left[\frac{-i}{2}(\dot\epsilon m e^{2\epsilon})
	x^2\right]\:,
	\label{can2}
	\xee
This leads to
	\xbea
	x\to x''&=&x,~~~~~p\to p''\:=\: p+me^{2\epsilon}\dot\epsilon x\;,
	\label{trans-xp'}\\
	H'\to H''&=&\frac{p^2}{2me^{2\epsilon}}+V(e^{\epsilon}x,t)+
	\frac{1}{2}\left[ \frac{d}{dt}(me^{2\epsilon}\dot\epsilon)-
	me^{2\epsilon}\dot\epsilon^2\right]\,x^2\;.
	\label{H'3}
	\xeea

\section{Time-Dependent Harmonic Oscillator}

The Hamiltonian of a time-dependent harmonic oscillator  with
mass $m=m(t)$ and frequency $\omega=\omega(t)$ is given by
	\xbe
	H=\frac{p^2}{2m(t)}+\frac{1}{2}\,m\omega(t)^2x^2\,.
	\label{star-h}
	\xee
For this system Eq.(\ref{H'3}) takes the form
	\xbe
	H''=\frac{p^2}{2me^{2\epsilon}}+
	\frac{1}{2}\left[\frac{d}{dt}(me^{2\epsilon}\dot\epsilon)
	+me^{2\epsilon}(\omega^2-\dot\epsilon^2)\right]x^2\;,
	\label{osc}
	\xee
where time-dependence of  $m,~\omega$ and $\epsilon$ are
suppressed for brevity.

Next let us choose $\epsilon(t)=\ln[m_0/m(t)]/2$ for a positive constant
$m_0$, so that $me^{2\epsilon}=m_0$. This choice yields
	\xbe
	H''=\frac{p^2}{2m_0}+\frac{1}{2}\,m_0\Omega^2x^2\;,
	\label{H'4}
	\xee
which is the Hamiltonian for a harmonic oscillator with constant mass 
$m_0$ and frequency
	\xbe
	\Omega:=\sqrt{\ddot\epsilon-\dot\epsilon^2+\omega^2}\:.
	\label{Omega}
	\xee
Requiring $\Omega$ to be independent of time, one can exactly solve
the Schr\"odinger equation for $H''$ which is now time-independent.
Using the canonical transformation defined by ${\cal U}'':=
({\cal U}'{\cal U})^\dagger$ and Eq.~(\ref{trans-u}), one then finds
the exact solution of  the Schr\"odinger equation for the original
harmonic oscillator. Thus the requirement $\Omega=\Omega_0=$ const.\
corresponds to a class of exactly solvable time-dependent harmonic
oscillators. Note that although time-dependent harmonic oscillators 
have been extensively studied during the past five decades
\cite{c-k} - \cite{pra}, for arbitrary choices of mass $m(t)$ and 
frequency $\omega(t)$ a closed expression for the time-evolution operator
in terms of $m(t)$ and $\omega(t)$ is not yet known. 

Next let us re-express the condition $\Omega=\Omega_0$ in terms of
$m$ and $\omega$. This leads to
	\xbe
	\ddot\epsilon-\dot\epsilon^2+\alpha^2=0\;,~~~~\alpha:=
	\sqrt{\omega^2-\Omega_0^2}\;,
	\label{condi}
	\xee
or alternatively
	\xbe
	\omega=\sqrt{\Omega_0^2+\frac{\ddot m}{2m}-
	\left(\frac{\dot m}{2m}\right)^2}\;.
	\label{condi-1}
	\xee
Hence, according to the above argument, the  time-dependent oscillators
whose mass $m$ and frequency $\omega$ satisfy (\ref{condi-1}) are
canonically equivalent to the time-independent
harmonic oscillator (\ref{H'4}) with $\Omega=\Omega_0$.

Next let us consider the case where the frequency $\omega$ is constant.
Then Eq.~(\ref{condi-1}) can be easily integrated to yield
	\xbe
	m(t)=m_0\left(\mu e^{\alpha t}+\nu e^{-\alpha t}\right)^2\;,
	\label{m=}
	\xee
where $\mu$ and $\nu$ are constants. Clearly, the Caldirola-Kanai
oscillator \cite{c-k} whose mass depends exponentially on time, i.e.,
$m=m_0e^{\gamma t}$ belongs to this class of oscillators. In fact,
Colegrave and Abdalla \cite{abd} have considered using the canonical
transformation (\ref{trans-xp}) to treat the oscillators with time-dependent
mass and fixed frequency, and in particular the Caldirola-Kanai oscillator. 
However, they perform the canonical transformation within the
classical context and then quantize the Hamiltonian. Fortunately the Hermiticity
requirement determines the quantum  Hamiltonian uniquely. Hence the lack of
knowledge about the precise unitary transformation corresponding to this
canonical transformation does not play much of a role in their analysis.

\section{Quantum to Classical Reduction of the Dynamical Equation and
the Ermankov Equation}

Consider the transformed harmonic oscillator Hamiltonian (\ref{osc}). If
the square bracket on the right hand side of  (\ref{osc}) vanishes, i.e.
	\xbe
	\frac{d}{dt}(me^{2\epsilon}\dot\epsilon) +me^{2\epsilon}(
	\omega^2-\dot\epsilon^2)=0\,,
	\label{I}
	\xee
then $H''$ describes a free particle with time-dependent mass. The
corresponding Schr\"odinger equation is exactly solvable. This means
that if one chooses $\epsilon$ such that the requirement (\ref{I})
is satisfied, then one obtains the solution of the Schr\"odinger equation
for the most general time-dependent harmonic oscillator.

Now if one introduces $\chi:=e^\epsilon$ and expresses Eq.~(\ref{I}) in
terms of $\chi$, one obtains
	\xbe
	\frac{d}{dt}(m\dot\chi)+m\omega^2\chi=0\;.
	\label{II}
	\xee
It is not difficult to recognize this equation as the classical equation
of motion for the time-dependent harmonic oscillator.

The reduction of the Schr\"odinger equation to the classical equation 
of motion is known since 1950's, \cite{jpn,ma-ma-tr,kim-}. It is 
nevertheless interesting to see that this reduction may be performed 
using a canonical transformation.  

It should be emphasized that any positive real solution of Eq.~(\ref{II})
may be used to obtain the exact solution of the Schr\"odinger
equation for the time-dependent harmonic oscillator (\ref{star-h}). 
If a positive solution of (\ref{II}) is found, then the time-evolution 
operator for the oscillator (\ref{star-h}) is given by
	\xbe
	U(t)={\cal U}(t)^\dagger{\cal U}(t)^{'\dagger}V(t)
	{\cal U}'(0){\cal U}(0)\;,
	\label{solution}
	\xee
where
	\xbea
	{\cal U}(t)&:=&e^{\frac{i}{2}(\ln\chi)\{x,p\}}\;,\xnn\\
	{\cal U}'(t)&:=&e^{-\frac{i}{2}(m\chi\dot\chi)x^2}\;,\xnn\\
	V(t)&:=&e^{-ia(t)p^2/2}\:,\xnn\\
	a(t)&:=&\int_0^t\frac{dt'}{m(t')\chi(t')^2}\;.\xnn
	\xeea

A slight generalization of condition (\ref{I}) is to demand that the
adiabatic approximation yields the exact solution of the Schr\"odinger
equation for the transformed Hamiltonian (\ref{osc}). In view of the
results reported in \cite{pra}, the condition of the exactness of
the adiabatic approximation for a time-dependent harmonic oscillator
is that the product of its mass and frequecy be a constant $k$. For the
oscillator (\ref{osc}), this means
	\xbe
	\left[\frac{d}{dt}(m e^{2\epsilon}\dot\epsilon)+
	e^{2\epsilon}(\omega^2-\dot\epsilon^2)\right]
	me^{2\epsilon}=k^2\;.
	\label{III}
	\xee
In terms of the variable $\chi:=e^{\epsilon}$,
Eq.~(\ref{III}) has the form
	\xbe
	\left[\frac{d}{dt}(m\dot\chi)+m\omega^2\chi\right]m\chi^{3}
	=k^2\;.
	\label{IV}
	\xee
For $k=0$, it reduces to 
Eq.~(\ref{II}). For $k\neq 0$, a simple rescaling of $\chi$
by $\sqrt{|k|}$, namely $\chi\to\chi':=\chi/\sqrt{|k|}$, leads to 
	\xbe
	\left[\frac{d}{dt}(m\dot\chi')+m\omega^2\chi'\right]
	m\chi^{'3}=\pm 1\;,
	\label{ax}
	\xee
where the minus sign corresponds to the case where the transformed
oscillator (\ref{osc}) has imaginary frequency. This means that
the relevant values of $k^2$ in (\ref{IV}) are $-1,~0$ and $1$. Again
any positive solution of any of the Eqs.~(\ref{ax}) leads to the
solution of the Schr\"odinger equation. The time-evolution operator
is still given by Eq.~(\ref{solution}), but now
	\[ V(t):=e^{-i\alpha(t)(p^2+k^2x^2)/2}\;.\]

Having  obtained the expression for the time-evolution operator,
one can easily compute the time-dependent position and momentum
operators. The result is
	\xbe
	x(t)=a(t) x + b(t) p\;,~~~~~~
	p(t)=c(t) x + d(t) p\;,
	\label{x-p=}
	\xee
where
	\xbea
	a&:=&\left(\frac{\chi}{\chi_0}\right)\left[
	\cos(k\alpha)-\frac{m_0\chi_0\dot\chi_0\sin(k\alpha)}{k}\right]\,,\xnn\\
	b&:=&\frac{\chi_0\chi\sin(k\alpha)}{k}\,,\xnn\\
	c&:=&\left(\frac{m\dot\chi}{\chi_0}-\frac{m_0\dot\chi_0}{\chi}\right)
	\cos(k\alpha)-\left(\frac{k}{\chi_0\chi}+\frac{m_0\dot\chi_0m\dot\chi}{k}\right)
	\sin(k\alpha)\,,\xnn\\
	d&:=&\left(\frac{\chi_0}{\chi}\right)\cos(k\alpha)+
	\left(\frac{\chi_0m\dot\chi}{k}\right)\sin(k\alpha)\,,\xnn\\
	m_0&:=&m(0)\,,~~~~\chi_0\::=\:\chi(0)\,,~~~~\dot\chi_0\::=
	\:\dot\chi(0)\,.
	\xnn	
	\xeea
The expressions corresponding to $k=0$ are obtained from (\ref{x-p=})
in the limit $k\to 0$. One can check that indeed $x(0)=x$, $p(0)=p$,
and $\left[x(t),p(t)\right]=i$.

It is remarkable to note that using the (quadratic) invariant theory of
Lewis and Riesenfeld \cite{le-ri}, Lewis \cite{lewis} had reduced the
solution of the Schr\"odinger equation for a harmonic oscillator with
time-dependent frequency to the solution of Eq.~(\ref{IV}) with $k=1$.
This equation was previously considered by Ermankov \cite{er}.
Here we obtained this equation by demanding that the adiabatic
approximation yields the exact solution of the Schr\"odinger equation for
the canonically transformed system. 

One may generalize the results of this section by performing other
time-dependent canonical transformations and demanding the result to
be exactly solvable. For example by requiring the transformed Hamiltonian
(\ref{osc}) to be one of the exactly solvable oscillators obtained in
Ref.~\cite{pra}, one obtains various generalizations of Eq.~(\ref{IV}).
These are, however, integro-differential equations whose solution seems
to be at least as difficult as Eq.~(\ref{IV}).

\section{Time-dependent Diffeomorphisms which Change the Metric}

The time-dependent dilatations which correspond to the choice $f(x)=x$
in (\ref{u}) form a very small class of  quantum canonical transformations
of the form (\ref{u}). As it is seen from Eqs.~(\ref{x}) and (\ref{p}) the
transformations induced on the position and momentum operators depend in
a complicated manner on $f(x)$. Some other choices of $f(x)$ for which 
these transformations can be calculated in a closed form are
	\xbe
	\begin{array}{cccc}
	f(x)=x^2 :&\left\{\begin{array}{ccc}
	x&\rightarrow& x'=\frac{x}{1-\epsilon(t) x}\\
	p&\rightarrow& p'=[1-\epsilon(t)x]\,p\,[1-\epsilon(t)x]
	\end{array}\right\}&{\rm for}&|\epsilon(t)x|<1\\
	&&&\\
	f(x)=e^{-\lambda x}:&\left\{\begin{array}{ccc}
	x&\rightarrow& x'=\frac{1}{\lambda}\ln\left[e^{\lambda  x}+
	\epsilon(t)\lambda \right]\\
	p&\rightarrow& p'=\sqrt{1+\frac{\epsilon(t)e^{\lambda x}}{\lambda}}
	\,p\,\sqrt{1+\frac{\epsilon(t)e^{\lambda x}}{\lambda }}
	\end{array}\right\}&{\rm for}&|\epsilon(t)\lambda|<1\;,
	\end{array}
	\label{eg}
	\xee
where $\lambda$ is a positive real number. As seen from these formulae,
the effect of these canonical transformations on the kinetic part $p^2/2m$
of the Hamiltonian is to make the mass $m$ also depend on the position.
This is precisely what happens when one considers a free particle moving
on a line with a metric $g$. In this case the quantum
Hamiltonian is given by \cite{super}
	\xbe
	{\cal H}= \frac{1}{2m}\left[ g^{-1/4}p\,g^{-1/2}p\,g^{-1/4}\right]\;.
	\label{curved}
	\xee
It is uniquely determined by the classical Hamiltonian ${\cal H}_c=
p^2/(2mg)$ and the self-adjointness requirement with respect to the
measure $\sqrt{g}\, dx$.

In view of  Eqs.~(\ref{curved}) and~(\ref{p}), one can easily infer the
fact that under the canonical transformations (\ref{u}) a free particle in
$\xR$ with the metric $g(x)=1$ is mapped to a free particle in $\xR$ with
a metric $g=[{\cal F}_2]^{-2}$. This is precisely the metric one would
obtain by performing the diffeomorphism $x\to x'={\cal F}_1$.
The converse of this statement is also true in the sense that for an arbitrary 
metric $g=g(x;t)$, there is a canonical transformation of the form (\ref{u}) 
which maps the problem to the ordinary one-dimensional free particle
problem provided that one can solve the pseudo-differential equation
	\[f(x)\,e^{\epsilon(t)f(x)\frac{d}{dx}}\,f^{-1}(x)=:{\cal F}_2(x)=
	[g({\cal F}_1(x);t)]^{-1/2}:=\left[g(e^{\epsilon(t)f(x)\frac{d}{dx}}
	\,x;t)\right]^{-1/2}\]
for $f(x)$. For the examples listed in (\ref{eg}), one has
	\xbea
	f=x^2~~~~   &\Longleftrightarrow& g=[1-\epsilon(t)x]^{-4}\;,\xnn\\
	f=e^{-\lambda x} &\Longleftrightarrow& g=\left[
	1+\frac{\epsilon(t)e^{\lambda x}}{\lambda }\right]^{-2}\;.
	\xnn
	\xeea

These considerations show that the one-dimensional quantum mechanics
of a free particle with position (and time) dependent mass is canonically
equivalent to that of a free particle with constant mass.

\section{Conclusion} 

In this article I have explored the quantum canonical transformations
corresponding to time-dependent diffeomorphisms of the configuration
space $\xR$. A special class of these transformations which are
associated with the time-dependent dilatations is used to obtain a new
class of exactly solvable time-dependent harmonic oscillators. A
well-known oscillator which belongs to this class is the Caldirola-Kanai
oscillator. 

Another application of time-dependent dilatations is in the reduction of the 
Schr\"odinger equation for the general time-dependent harmonic oscillator
to the corresponding classical equation of motion. Although the relation
between the quantum and classical dynamical equations has been well-known,
its direct realization via time-dependent dilatations is a new result. 
More specifically, I have shown that
	\begin{itemize}
	\item[--] if one uses a positive solution $\chi= \chi(t)$ of
the classical equation of motion (\ref{II}) to perform the time-dependent
quantum dilatation $x\to \chi(t)x$, then one obtains the solution of the 
Schr\"odinger equation;
	\item[--] if one performs a quantum dilatation $x\to\chi(t)x$, 
canonically transforms the resulting generalized harmonic oscillator 
Hamiltonian to a Hamiltonian of the standard form, and requires the 
adiabatic approximation to yield the exact solution of the Schr\"odinger 
equation for the latter Hamiltonian, then one obtains a second order 
differential equation in the dilatation parameter $\chi$ which involves 
a parameter $k^2=-1,0,1$. For $k=0$, this is the classical equation of 
motion. For $k=1$, this is known as the Ermankov equation which
is also obtained by applying the Lewis-Riesenfeld invariant theory to the 
time-dependent harmonic oscillator.
	\end{itemize}
I have also briefly commented on the quantum canonical transformations
which correspond to metric-changing diffeomorphisms. 

The direct generalization of the analysis presented in this article to higher
dimensions is not difficult.  In fact, the $n$-dimensional analogue of
(\ref{u}), i.e.,
	\xbe
	{\cal U}:=\exp\left[ \frac{i\epsilon(t)}{2}\sum_{i=1}^n\{f^i(x),p_i\}
	\right]=\exp \left[i\epsilon(t)\sum_{i=1}^n\sqrt{f^i(x)}\,p_i\,
	\sqrt{f^i(x)}\right]\;,
	\label{u-n}
	\xee
corresponds to diffeomorphisms of $\xR^n$. Here $x=(x^1,\cdots,x^2)$
and $f^i(x)$ are smooth real-valued functions of $x$. For $n>1$ there 
are metrics on $\xR^n$ which are not related to the Euclidean metric 
by a diffeomorphism. Thus, in general canonical transformations of the 
form (\ref{u-n}) do not relate the dynamics of a free particle on an 
arbitrary curved $\xR^n$ to that of $\xR^n$ with Euclidean geometry.

\section*{Acknowledgments}
Parts of this project were completed while I was a member of
the Theoretical Physics Institute of the University of Alberta.
I wish to acknowledge the financial support of the Killam
Foundation of Canada and thank Drs.~M.~Razavy and B.~Darian
for interesting discussions.
\vspace{.1cm}

{\bf  Note:} After the completion of this article Refs.~\cite{razavy,wa-oh-kw}
were brought to my attention.  In Ref.~\cite{razavy} the author uses the 
quantum dilatations (\ref{trans-xp}) to study the solution of the Schr\"odinger
equation with a moving boundary condition. Ref.~\cite{wa-oh-kw} gives
a systematic analysis of the most general exponential quadratic operator
in $x$ and $p$, whose special case is the quantum dilatations studied
in this article.


\end{document}